\documentclass[11pt]{article}

\usepackage{tikz}
\usepackage{amssymb,amsmath,amsfonts}
\usepackage{latexsym,bbold}
\usepackage[totalwidth=17cm,totalheight=23.8cm]{geometry}

\newcommand{\D}{{\partial\!\!\!\slash}}
\newcommand{\cD}{{\mathcal D}}
\newcommand{\1}{{\mathbb{1}}}

\newcommand\be{\begin{eqnarray}}
\newcommand\ee{\end{eqnarray}}

\begin{document}

\title{One-loop $\beta$-function for an infinite-parameter family of gauge theories}
\author{Kirill Krasnov \\ \it{School of Mathematical Sciences, University of Nottingham}\\ \it{University Park, Nottingham, NG7 2RD, UK}}

\date{January 2015}
\maketitle

\begin{abstract} We continue to study an infinite-parametric family of gauge theories with an arbitrary function of the self-dual part of the field strength as the Lagrangian. The arising one-loop divergences are computed using the background field method. We show that they can all be absorbed by a local redefinition of the gauge field, as well as multiplicative renormalisations of the couplings. Thus, this family of theories is one-loop renormalisable. The infinite set of $\beta$-functions for the couplings is compactly stored in a renormalisation group flow for a single function of the curvature. The flow is obtained explicitly. 
\end{abstract}

\section{Introduction} 

The purpose of this paper is to compute the one-loop beta-function for an infinite-parametric family of gauge theories introduced in \cite{CFK}. The Lagrangian of these theories is given by an arbitrary function of the {\it self-dual} part of the field strength. In paper \cite{CFK} we have performed a perturbative study of these theories, and analysed the arising scattering amplitudes. We have also computed the beta-function for the first non-trivial coupling describing deformations away from Yang-Mills. It was noted that this family of theories can be expected to be renormalisable at the one-loop level, by a local redefinition of the gauge field, as well as by multiplicative renormalisation of the coupling constants. In this paper we use the background field method to compute the arising divergences, and establish the one-loop renormalisability. We also extract the full one-loop beta-function for an infinite number of coupling constants parameterising our theories. 

Given that the calculation that leads to the $\beta$-function is quite technical, it is worth stating the result already in the Introduction. The family of theories we study is described by the following Lagrangian
\be\label{L-conn}
{\cal L} = M^4 f(F^a_{MN}/M^2).
\ee
Here $M$ is a mass scale introduced so as to make all couplings dimensionless. The quantity $F^a_{MN}$ is the self-dual part of the field strength for the gauge field, $a$ is the Lie algebra index, and $M,N$ are unprimed spinor indices, see below for our spinor index conventions. The function $f$ is an arbitrary gauge- and Lorentz-invariant function of $F^a_{MN}$. In case this function is analytic near zero, it can be expanded in powers of the argument. The first non-trivial term is then the YM Lagrangian, and the next term, cubic in the field strength, describes the first non-trivial chiral "deformation" of YM theory. It is clear that a theory of type (\ref{L-conn}) is specified by an infinite number of coupling constants, coefficients of the power series expansion of the function $f$. Apart from the YM term quadratic in the curvature, all other interactions described by (\ref{L-conn}) are power-counting non-renormalisable and contain inverse powers of $M^2$.

We compute one-loop divergences using the background field method. We work in Riemannian signature. The one-loop renormalisability of (\ref{L-conn}) means that, after a local field redefinition, the remaining one-loop divergences are taken care of by counter terms of the type already contained in (\ref{L-conn}). Thus, the one-loop running of the whole infinite set of the coupling constants can be encoded as the running of the dimensionless function $f$
\be
\frac{\partial f(x)}{\partial \log\mu} = \beta_f(x),
\ee
where $\beta_f$ is some gauge- and Lorentz-invariant function of the dimensionless self-dual part of the curvature $x^a_{AB}:=F^a_{AB}/M^2$. The result of our calculation is
\be\label{beta-f}
\beta_f(x) =\frac{1}{(4\pi)^2} \frac{1}{6} \Big[  x^{ab}_{AB} x^{ba AB}  -3 ((f'')^{-1})^{ab}_{AB}{}^{AB}  (f')^{bc}_{MN} x^{ca MN} \\ \nonumber
+ 3 ((f'')^{-1})^{ab}_{AM}{}^{BN} (f')^{bc}_{B}{}^C ((f'')^{-1})^{cd}_{CN}{}^{DM}(f')^{da}_D{}^A  \Big].
\ee
Here $(f')^{a AB}$ and $(f'')^{ab AB CD}$ are the matrices of the first and second derivatives of the function $f$
\be\label{f-ders}
(f')^{a AB}:= \frac{\partial f}{\partial x^a_{AB}}, \qquad (f'')^{ab AB CD} := \frac{\partial^2 f}{\partial x^a_{AB} \partial x^b_{CD}}, 
\ee
and $(f'')^{-1}$ is the matrix inverse to $f''$. We also use the notation 
$$x^{ab}_{AB}:= C^{aeb} x^e_{AB}, \qquad (f')^{ab}_{AB}:= C^{aeb} (f')^e_{AB},$$ 
where $C^{abc}$ are the Lie algebra structure constants.

Note that the result (\ref{beta-f}) is a homogeneity degree zero function in $f$, as it should be. As a further check, we note that for Yang-Mills 
\be
f_{\rm YM}(x) = \frac{1}{4g^2} (x^a_{AB})^2,
\ee
and thus
\be\nonumber
(f'_{\rm YM})^{ab}_{AB} = \frac{1}{2g^2} x^{ab}_{AB}, \qquad ((f''_{\rm YM})^{-1})^{ab}_{ABCD} = 2g^2 \delta^{ab} \epsilon_{A(C}\epsilon_{|B|D)}.
\ee
A simple computation then gives
\be
\beta^{\rm YM}_f = \frac{11 C_2}{6(4\pi)^2} (x^a_{AB})^2,
\ee
where $C_2$ is the quadratic Casimir. This gives the correct running of $1/4g^2$. 

When both the Lagrangian (\ref{L-conn}) and the $\beta$-function (\ref{beta-f}) are expanded in powers of $x^a_{AB}$, one can read off an infinite set of beta-functions for the couplings stored in $f$. In the main text we carry out this exercise for the first non-trivial coupling parameterising the $F^3$ interaction. We reproduce the result obtained in \cite{CFK} by a different method. Some further comments on the interpretation of (\ref{beta-f}) are contained in the Discussion section.

Let us give an outline of how (\ref{beta-f}) is computed. All details are given in the main text. The straightforward application of the background field method to (\ref{L-conn}) runs into a difficulty. The problem is that the second-order operator that arises by linearising (\ref{L-conn}) is not of Laplace-type, even after gauge-fixing. We alleviate the problem by passing to the first-order formulation, by "integrating in" an auxiliary field. The linearisation of the resulting Lagrangian then gives a first-order operator that turns out to be of Dirac-type. Its square is of Laplace-type, which makes the well-developed heat-kernel technology for operators of such type applicable. This allows for a straightforward, even thought at times technical, computation of the divergent parts of the relevant regularised determinant. 

The organisation of this paper is as follows. In the next Section we reduce the problem to a calculation of a certain determinant, which is then analysed in Section \ref{sec:det}. We apply the result obtained to the problem at hand in Section \ref{sec:beta}. This is where the result (\ref{beta-f}) is obtained. An application to a 2-parameter family of theories is considered in Section \ref{sec:appl}, where the result obtained in \cite{CFK} is reproduced. We conclude with a discussion.

\section{Setup of the calculation}
\label{sec:setup}

\subsection{Non-Laplace type operator}

Before we start the calculation, let us explain why we need to do more than just apply the standard background field method to (\ref{L-conn}). The problem with the straightforward application of the method is that the linearisation of the Lagrangian (\ref{L-conn}) around an arbitrary background 
\be\label{L-kin}
{\cal L}^{(2)} = \frac{1}{2} (f'')^{ab ABCD} (\partial a)^a_{AB} (d a)^b_{CD} + M^2 (f')^{a AB} [[a,a]]^a_{AB}
\ee
gives a non-Laplace type operator. Here $a^a_{AB}$ is the gauge field perturbation, and the Lagrangian is expanded to second order. The matrices $f', f''$ are the already encountered (\ref{f-ders}) matrices of the first and second derivatives of the function $f$. The other quantities are defined as follows
\be
(d a)^a_{AB} := 2d_{(AA'} a^a_{B)}{}^{A'}, \qquad [[a,a]]^a_{AB} :=  C^{abc} a^b_{AA'} a^c_B{}^{A'}.
\ee
Here $d$ is the covariant derivative with respect to the background gauge field. It is clear that the second order differential operator appearing in (\ref{L-kin}) is not of the Laplace type because the indices of the derivatives $\partial_{AA'}$ contract either with a spinor index of the gauge field perturbation or with a spinor index of the matrix of second derivatives. Thus, we have a second-order operator $\Delta$ acting on Lie algebra-valued one-forms $a_\mu^a$ with the leading second order part of the form 
\be
(\Delta a)^a_\mu = M^{ab \mu\nu\rho\sigma} \partial_\nu \partial_\rho a_\sigma^b+\ldots,
\ee
for a certain matrix $M^{ab \mu\nu\rho\sigma}$. This matrix is not proportional to $\eta^{\nu\rho}$, and thus the operator does not reduce to a Laplace-type one with the leading part $M^{ab\mu\sigma} \partial^2 a_\sigma^b$ for some matrix $M^{ab\mu\sigma}$. Thus, the well-developed heat-kernel technology for Laplace-type operators is not applicable in our case, at least not directly. 

\subsection{The first-order formulation}

To apply the heat-kernel technology to (\ref{L-kin})  we would need to develop the associated expansion for non-Laplace type operators, which is possible, but not easy. We will choose another route. Thus, we will first reformulate our theories in the first-order form by "integrating in" an auxiliary field. The first order operator arising in this context turns out to be of Dirac-type, with square given by an operator of Laplace-type. This will reduce the problem at hand to the well-studied Laplace-type operators. 

Let us consider the following action principle
\be\label{action}
S[A,B] = \int B^{a MN} F^a_{MN} - M^4 V(B/M^2),
\ee
where $F^a_{MN}$ is the self-dual part of the curvature of $A^a_{MM'}$, and $B^a_{MN}$ is an auxiliary field. The field equations that follow from (\ref{action}) are
\be\label{feqs}
d_{A'}{}^E B^a_{AE} = 0, \qquad F^a_{MN}/M^2 =  (V')^{a}_{MN},
\ee
where $V^{(1) a}_{MN}$ is the matrix of first derivatives of the potential function $V(B)$
\be
(V')^{a MN} = \frac{\partial V}{\partial (B^{a MN}/M^2)}.
\ee
We can solve the second field equation for $B/M^2$ in terms of $F/M^2$, and substitute the result to (\ref{action}) to obtain a theory of type (\ref{L-conn}). This identifies $V(B/M^2)$ as the Legendre transform of the function $f(F/M^2)$. So, we have an equivalent formulation, at least classically. One can also check explicitly that the exact equivalence continues to hold at the one-loop level, see below. 

The usual Yang-Mills theory can be described in this language, and corresponds to the choice
\be
M^4 V_{YM}(B/M^2) = g^2 (B^a_{MN})^2.
\ee
Indeed, in this case, integrating the auxiliary field $B$ out we get $B=(1/2g^2) F$ and
\be
S_{\rm YM}[A] = \frac{1}{4g^2} \int (F^a_{MN})^2,
\ee
which is the usual Euclidean signature YM action. 

In general, field equations (\ref{feqs}) imply a second-order differential equation for the connection. First, one solves the second equation for $B^a_{MN}$ as a function of $F^a_{MN}$. This is an algebraic task. In the second step one substitutes the solution into the first equation, and obtains a second order PDE for $A^a_{MM'}$. It is easy to see that this is the same PDE that is obtained directly from the Lagrangian (\ref{L-conn}). 

\subsection{Linearisation}

We now want to compute the one-loop $\beta$-function for (\ref{action}) using the background field method. As we shall see, the associated problem can be reduced to a Laplace-type operator. We first linearise the action (\ref{action}) around an arbitrary on-shell configuration of fields $B,A$. We denote the fluctuations by $b,a$. The resulting quadratic in the fluctuations Lagrangian is
\be\label{L-ab}
{\cal L}^{(2)} = 2 b^{a NM} d_{MM'} a^a_N{}^{M'} + C^{abc} B^{a MN} a^b_{MM'} a^c_N{}^{M'} - \frac{1}{2} (V'')^{ab}_{MN PQ} b^{a MN} b^{b PQ}.
\ee
Here $d_{AA'}$ is the covariant derivative with respect to the background connection, and the matrix $(V'')^{ab}_{MN PQ}$ is that of second derivatives of the potential function $V(B)$. The reason why in the first term we wrote $b^{a NM}$ with the order $NM$ of the spinor indices will become clear below. 

Before we proceed with our analysis, let us check that the functional determinant arising from (\ref{L-ab}) is the same one that arises from (\ref{L-kin}). Thus, let us integrate out the perturbation of the auxiliary field. The field equation for $b^a_{MN}$ reads
\be\label{b-eqn}
(d a)^a_{MN} = (V'')^{ab}_{MNPQ} b^{b PQ}.
\ee
Given that the function $V$ is the Legendre transform of $f$ we have
\be
((V'')^{-1})^{ab}_{MNPQ} = (f'')^{ab}_{MNPQ}, 
\ee
and thus the solution of (\ref{b-eqn}) is
\be
b^{a}_{MN} = (f'')^{ab}_{MNPQ} (da)^{b PQ}.
\ee
Using the background field equation $B^a_{MN}=(f')^a_{MN}$, we then see that (\ref{L-ab}) with $b^a_{MN}$ integrated out is precisely (\ref{L-kin}). This establishes the one-loop equivalence of the first- and second-order formulations. 

\subsection{Gauge-fixing}

The kinetic term in (\ref{L-ab}) is degenerate, as it is built from the operator that sends the linearised connection field $a^a_{AA'}$ with its $4\times {\rm dim}({\mathfrak g})$ components into an object $d_{M'(M} a^a_{N)}{}^{M'}$ with $3\times {\rm dim}({\mathfrak g})$ components. We thus need to gauge fix in order to be able to invert the kinetic term. 

The gauge-fixing fermion can be taken to be the usual one in the context of Yang-Mills theories
\be\label{psi}
\Psi = 2 \bar{c}^a d^{M}_{M'} a^a_M{}^{M'} - 2 M^{ab} \bar{c}^a h_c^b,
\ee
where $\bar{c}^a$ is an anti-ghost, $h_c^a$ is an auxiliary field, and we have allowed ourselves to introduce some matrix $M^{ab}$ instead of the gauge-fixing parameter. This can, in principle, depend on the background field $B^a_{MN}$. 

The BRST variation of (\ref{psi}) is
\be\label{Lg}
s \Psi = 2 \bar{c}^a d^{M}_{M'} d_M{}^{M'} c^a + 2 h_c^a d^{M}_{M'} a^a_M{}^{M'}- 2 M^{ab} h_c^a h_c^b.
\ee
The first term here gives the kinetic term for the ghosts, while the second term should be added to the bosonic part of the Lagrangian, and removes the degeneracy discussed above.

A particularly convenient way to combine the last two terms in (\ref{Lg}) with (\ref{L-ab}) is to introduce a new component to the $b^a_{MN}$ field
\be
\tilde{b}^{a NM} = b^{a NM} + \epsilon^{NM} h_c^a.
\ee
The new field $\tilde{b}^{a NM}$ is no longer $MN$ symmetric, and the two indices become distinct. This explains why we wrote $b^{a NM}$ in the kinetic term in (\ref{L-ab}). With the new field introduced, the first term in (\ref{L-ab}) plus the second term in (\ref{Lg}) become simply $2 \tilde{b}^{a NM} d_{MM'} a^a_N{}^{M'}$. Further, he last term in (\ref{Lg}) combines with the last term in (\ref{L-ab}) by introducing
\be
(\tilde{V}'')^{ab}_{MN PQ}=(V'')^{ab}_{MN PQ} - M^{ab} \epsilon_{MN} \epsilon_{PQ}.
\ee
The full gauge-fixed Lagrangian then becomes
\be\label{L2'}
{\cal L}^{(2)} = 2 \tilde{b}^{a NM} d_{MM'} a^a_N{}^{M'} + f^{abc} B^{a MN} a^b_{MM'} a^c_N{}^{M'} - \frac{1}{2} (\tilde{V}'')^{ab}_{MN PQ} \tilde{b}^{a MN} \tilde{b}^{b PQ},
\ee
which is the same as (\ref{L-ab}), but now for the field $\tilde{b}$. From now on we will omit the tilde from the $b$-field for brevity. 

\subsection{The operator arising}

The operator in (\ref{L2'}) is of the type 
\be\label{D}
D = \left( \begin{array}{cc} A & \D \\ \D^* & B  \end{array} \right),
\ee
where 
$$ \D : V\otimes S_- \to V\otimes S_+, \quad \D^* : V\otimes S_+\to V\otimes S_-$$
are the usual chiral Dirac operators and
$$A\in {\rm End}(V\otimes S_+), \quad B \in {\rm End}(V\otimes S_-).$$
Here $S_\pm$ are the spaces of unprimed and primed spinors. In the case of (\ref{L2'}) the bundle $V$ is given by
\be\label{bundle}
V= {\mathfrak g}\otimes S_+ \, ,
\ee
and it is the first index $N$ of $b^{a NM}$ that is considered as taking values in the $S_+$ part of (\ref{bundle}). Further, in our case the endomorphism $A$ is simply
\be\label{A}
A_{AB} \equiv A^{ab}_{MA NB} = -\frac{1}{2} (\tilde{V}'')^{ab}_{MA NB}.
\ee
Finally, the endomorphism $B$ is 
\be
B_{A'B'}\equiv B^{ab}_{MN} \epsilon_{A'B'} = C^{aeb} B^e_{MN}\epsilon_{A'B'}.
\ee
This can be written in more abstract notations as
\be\label{form-B}
{\rm End}(V\otimes S_-)\ni B = \tilde{B}\cdot \1_{S_-},\qquad B\in {\rm End}(V),
\ee
with $\1_{S_-}$ being the identity operator on $S_-$. In our case 
\be\label{tB}
\tilde{B} \equiv \tilde{B}^{ab}_{MN} = C^{aeb} B^e_{MN}.
\ee
We now proceed with the calculation of the (divergent part of) the determinant of (\ref{D}), and return to the problem at hand afterwards. 

\section{Calculation of the determinant}
\label{sec:det}

\subsection{Problem setup}

Consider the Dirac-type operator (\ref{D}), with $A, B$ being some symmetric endomorphisms, so that the operator is self-adjoint. We assume that $B$ is of the form (\ref{form-B}). This operator acts on columns $$\left( \begin{array}{c} b \\ a \end{array}\right), \qquad b\in V\otimes S_+, \quad a\in V\otimes S_-,$$ where $V$ is some vector space. In other words, the operator (\ref{D}) acts on $V$-valued Dirac spinors, and it is assumed that the chiral Dirac operators $\D,\D^*$ know how to act on $V$-valued sections.

We would like to compute the divergent parts of the regularised determinant of the operator (\ref{D}). One way to do this computation would be to eliminate either $a$ or $b$ fields, and consider the resulting second-order operator. We already know that if we integrate out $b$ we get a non-Laplace type operator. It is easy to check that what one obtains is exactly the operator in (\ref{L-kin}). Thus, eliminating $b$ is not a good idea. Another possibility is to eliminate $a$, and obtain a second order differential operator for $b$. It is easy to see that this has a better chance to work, because the arising operator is of Laplace type. However, in the process of integrating out $a$ one needs to invert the matrix $B$, which is possible, but cumbersome. 

Instead of trying to reduce the problem at hand to a second-order Laplace type operator by integrating one of the two fields $a,b$, let us consider the square
\be
D^2 = \left( \begin{array}{cc} \D\D^* + A^2 & A\D + \D B \\ B \D^*+\D^* A & \D^* \D+B^2 \end{array} \right).
\ee
It is clear that the arising second order operator is of Laplace type, and hence the divergences of interest are easily computed using the heat-kernel techniques. 

To perform the computation we will need to rewrite $D^2$ as a multiple of 
\be\label{type}
-\cD^\mu \cD_\mu + E,
\ee
where $\cD$ is some covariant derivative and $E$ is some endomorphism. We will compute both $\cD$ and $E$ in the following subsections.

\subsection{Spinor index conventions and Lichnerowicz formuli}

To do calculations with (\ref{D}) we will use spinor index notations. Thus, our conventions for the actions of $\D,\D^*$ are as follows. Let $\lambda_A\in S_+, \lambda_{A'}\in S_-$. Then
$$(\D\lambda')_A = d_{AA'} \lambda^{A'}, \qquad (\D^*\lambda)_{A'} = d_{A'}{}^A \lambda_A,$$
where $d_{AA'}$ is the usual covariant derivative operator in spinor notations. 

Let us use the spinor index notations to compute the squares $\D\D^*$ and $\D^* \D$. We expect these to be multiples of the corresponding Laplacians plus curvature terms, so let us work out what appears. We have
\be\label{dir2}
(\D\D^* \lambda)_A = d_{AA'} d^{A'E}\lambda_E = - \frac{1}{2} d^{M}{}_{M'} d_M{}^{M'} \lambda_A + \frac{1}{2} F_A{}^E\lambda_E, \\ \nonumber
(\D^*\D \lambda')_{A'} = d_{A'}{}^E d_{EE'} \lambda^{E'} = - \frac{1}{2} d^{M}{}_{M'} d_M{}^{M'} \lambda_{A'} +\frac{1}{2} F_{A'E'} \lambda^{E'}.
\ee
Here $F_{AE},F_{A'E'}$ are the self and anti-self-dual parts of the curvature of $d_\mu$, defined by the following formula
$$2d_{[\mu} d_{\nu]} \lambda = F_{\mu\nu}\lambda,$$
and the same for the action on $\lambda'$. Then the following decomposition of $F_{\mu\nu}\in {\rm End}(V)$ is valid
\be\label{F-sa}
F_{MM'NN'} = \frac{1}{2} F_{MN}\epsilon_{M'N'} + \frac{1}{2}F_{M'N'}\epsilon_{NM},
\ee
so that
$$F_{MN}\equiv F_{MM'N}{}^{M'}, \qquad F_{M'N'} \equiv F^M{}_{M'MN'}.$$
Equalities (\ref{dir2}) show that $2D^2$ is an operator of type (\ref{type}). 

\subsection{Computation of the connection}

To compute the connection that makes $2D^2$ into an operator (\ref{type}) we first rewrite the off-diagonal elements in a suggestive form
\be\label{d2-1}
2D^2=\left( \begin{array}{cc} -d^\mu d_\mu +F + 2A^2 & -2C^\mu d_\mu + 2(\D B) \\ -2C^{*\mu}d_\mu +2(\D^* A) & -d^\mu d_\mu +F'+2B^2 \end{array} \right).
\ee
Here
$$ F\in{\rm End}(V\otimes S_+), \qquad F'\in {\rm End}(V\otimes S_-)$$
are the self- and anti-self-dual parts of the curvature viewed as operators, and
$$C_\mu: V\otimes S_-\to V\otimes S_+, \qquad C_\mu^*: V\otimes S_+\to V\otimes S_-$$
are operators acting as
$$ (C_\mu \lambda')^{A} = C_\mu{}^A{}_{A'}\lambda^{A'}, \qquad
(C^*_\mu \lambda)_{A'} = C^*_\mu{}^A{}_{A'} \lambda_A.$$
The components of $C_\mu,C^*_\mu$ are given by
\be
C^M{}_{M'}{}^A{}_{A'} = \epsilon_{A'M'}(A+\tilde{B}\1_{S_+})^{AM}, \qquad
C^{*M}{}_{M'}{}^A{}_{A'} = \epsilon_{M'A'}(A+\tilde{B}\1_{S_+})^{MA}.
\ee
Finally, the derivatives of $B,A$ appearing in (\ref{d2-1}) are as follows
$$(\D B) : V\otimes S_-\to V\otimes S_+, \qquad (\D^* A): V\otimes S_+\to V\otimes S_-,$$
with the explicit expressions being
$$ ((\D B)\lambda')_{A} \equiv  (d_{AA'} \tilde{B}) \lambda^{A'}, \qquad
((\D^* A)\lambda)_{A'} \equiv (d_{A'}{}^A A_{A}{}^E) \lambda_E.$$
This immediately tells us that
\be
\cD_\mu = \left( \begin{array}{cc} d_\mu & C_\mu \\ C_\mu^* & d_\mu \end{array} \right)
\ee
and
\be
E = \left( \begin{array}{cc} F+2A^2 +C^\mu C_\mu^* & 2(\D B)+(d^\mu C_\mu) \\ 2(\D^* A) +(d^\mu C_\mu^*) & F'+2B^2 +C_\mu^* C^\mu \end{array} \right).
\ee
After some simplifications the endomorphism becomes
\be
E = \left( \begin{array}{cc} F-2(A \tilde{B}+\tilde{B} A) - 2\tilde{B}^2\cdot \1_{S_+}  & (\D (B-A))\\ (\D^* (A-B))  & F'-({\rm Tr}(A^2)+{\rm Tr}(A)\tilde{B}+\tilde{B}\,{\rm Tr}(A))\cdot \1_{S_-} \end{array} \right),
\ee
where we introduced new maps
$$
(\D (B-A) \lambda')_A \equiv (d_{MA'} (\tilde{B}\cdot \1_{S_+} - A)_A{}^M)\lambda^{A'}, \qquad
(\D^* (A-B) \lambda)_{A'} \equiv (d_{A'}{}^A (A-\tilde{B}\cdot \1_{S_+})_{A}{}^E ) \lambda_E.$$

\subsection{Curvature}

The curvature of $\cD_\mu$ is given by
\be
{\cal F}_{\mu\nu} = 2\cD_{[\mu} \cD_{\nu]} = \left( \begin{array}{cc} F_{\mu\nu} + C_\mu C^*_\nu - C_\nu C^*_\mu & 2d_{[\mu} C_{\nu]} \\ 2 d_{[\mu} C^*_{\nu]} & F_{\mu\nu} + C^*_\mu C_\nu - C^*_\nu C_\mu \end{array}\right).
\ee

\subsection{The heat-kernel formula}

The quantity we want to compute is, see e.g. \cite{Vassilevich:2003xt}
\be
a_4 = \frac{1}{(4\pi)^2} \frac{1}{12} \int {\rm Tr} \left( {\cal F}_{\mu\nu} {\cal F}^{\mu\nu} + 6 E^2\right).
\ee

\subsection{Off-diagonal terms}

First derivatives of the matrices $A,\tilde{B}$ enter only via the off-diagonal terms. These off-diagonal terms do not mix with the diagonal when the trace is taken. It is thus pertinent to analyse these clearly separable derivative terms first. 

Consider the contribution from the term $6E^2$ first. Squaring the off-diagonal term produces two diagonal terms that become the same under the trace in $V$. Thus, we can write for this contribution
\be\label{off-1}
12 \,{\rm Tr}_V \left( d^M{}_{A'}(A-\tilde{B}\cdot \1_{S_+})_{AM} d^{NA'} (A-\tilde{B}\cdot \1_{S_+})_N{}^A\right).
\ee
The other contribution is from ${\cal F}^2$. Again, there are two equal terms and we get
\be\label{off-2}
8\,{\rm Tr}_V \left( d_{[\mu} C_{\nu]}{}_{AA'} d^{[\mu} C^{*\nu]}{}^{AA'}\right).
\ee
While (\ref{off-1}) and (\ref{off-2}) are both quadratic in first derivatives of $A,B$, the space-time indices are contracted differently. We would like to bring the expression (\ref{off-2}) into the form similar to (\ref{off-1}), plus contributions without derivatives. This is achieved by a sequence of identities.

The idea is first to rewrite the contraction of two anti-symmetric tensors in (\ref{off-2}) as the contraction of the corresponding self-dual parts, plus a term containing $\epsilon_{\mu\nu\rho\sigma}$. The necessary identity is
\be\label{off-3}
2 d_{[\mu} C_{\nu]} d^{[\mu} C^{*\nu]} = 2 (d_{M'(M} C_{N)}{}^{M'}) (d_{N'}^{(M} C^{*N)N'}) - \epsilon^{\mu\nu\rho\sigma} d_\mu C_\nu d_\rho C^*_\sigma,
\ee
where we suppressed the spinor indices of $C_\mu,C_\mu^*$ for brevity. We can integrate by parts in the last term, and modulo a total derivative write
\be\label{off-4}
\epsilon^{\mu\nu\rho\sigma} d_\mu C_\nu d_\rho C^*_\sigma = \frac{1}{2}\epsilon^{\mu\nu\rho\sigma} C_\rho F_{\mu\nu} C^*_\sigma = \frac{1}{2} C^M_{M'} F_{MN} \circ C^{*NM'} - \frac{1}{2} C^{MM'} F_{M'N'} \circ C_M^{*N'}.
\ee
Note that here the curvature acts on the quantity $C^*$ that takes values in ${\rm End}(V)$, not in $V$. To get this identity we have used the following spinor expression for the 4-dimensional $\epsilon$-tensor
\be
\epsilon^{\mu\nu\rho\sigma} = \epsilon^{M'N'}\epsilon^{R'S'} \epsilon^{MR}\epsilon^{NS}- \epsilon^{M'R'}\epsilon^{N'S'}\epsilon^{MN}\epsilon^{RS},
\ee 
as well as (\ref{F-sa}). In our case the last term in (\ref{off-4}) does not contribute because both $C_\mu,C^*_\mu$ are proportional to the primed spinor metric, which contracting with the symmetric $F_{M'N'}$ gives zero. Thus, reintroducing the spinor indices on $C_\mu^{AA'},C_\mu^{*AA'}$ we get for (\ref{off-2})
\be\label{off-5}
- 8\,{\rm Tr}_V\left( d_{A'(M} X_{AN)} d^{A'(M} X^{N)A}\right)
+ 4\,{\rm Tr}_V\left( X_A{}^M F_{MN} \circ X^{NA}\right),
\ee
where we denoted
\be\label{X}
X_{AB}\equiv (A+\tilde{B}\cdot \1_{S_+})_{AB},
\ee
and in the last term the curvature acts on an object in ${\rm End}(V\otimes S_+)$, which is denoted by the $\circ$ symbol.

The final identity that we need is one reducing (\ref{off-1}) to an expression of the type appearing in (\ref{off-5}). This is again achieved by an integration by parts. We have
\be\label{off-6}
d_{A'}^M X_{AM} d^{A'N} X_N{}^A = - X_{AM} d_{A'}^M d^{A'N} X_N{}^A = - X_{AM} d^{A'N} d_{A'}^M  X_N{}^A - X_{AM} F^{MN} \circ X_N{}^A \\ \nonumber
= d_{A'N} X_{AM} d^{A'M}  X^{NA} - X_{AM} F^{MN} \circ X_N{}^A.
\ee
But then we have
\be
 d_{A'N} X_{AM} =  d_{A'(M} X_{AN)} - \frac{1}{2} \epsilon_{NM} d_{A'}^E X_{AE}.
 \ee
 Substituting this into (\ref{off-6}) we get
 \be
 d_{A'}^M X_{AM} d^{A'N} X_N{}^A = d_{A'(M} X_{AN)} d^{A'(M}  X^{N)A} -\frac{1}{2} d_{A'}^M X_{AM} d^{A'N} X_N{}^A  - X_{AM} F^{MN} \circ X_N{}^A
 \ee
 and thus
 \be
 d_{A'(M} X_{AN)} d^{A'(M}  X^{N)A}=\frac{3}{2} d_{A'}^M X_{AM} d^{A'N} X_N{}^A +X_{AM} F^{MN} \circ X_N{}^A,
 \ee
 which is the identity we need. 

Collecting all the above, we see that the total contribution of the off-diagonal terms is given by
\be
12 \,{\rm Tr}_V \left( d^M{}_{A'}(A-\tilde{B}\cdot \1_{S_+})_{AM} d^{NA'} (A-\tilde{B}\cdot \1_{S_+})_N{}^A\right) \\ \nonumber
-12 \,{\rm Tr}_V \left( d^M{}_{A'}(A+\tilde{B}\cdot \1_{S_+})_{AM} d^{NA'} (A+\tilde{B}\cdot \1_{S_+})_N{}^A\right) \\ \nonumber
-4\,{\rm Tr}_V\left( X_A{}^M F_{MN} \circ X^{NA}\right),
\ee
where we kept the matrix (\ref{X}) in the last term. Thus, in the terms containing the derivatives of $A,\tilde{B}$ only the cross-term remains and we get for the contribution from the off-diagonal terms
\be\label{off*}
24\, {\rm Tr}_V\left( d_{AA'} {\rm Tr}(A) d^{AA'} \tilde{B}\right) -4\,{\rm Tr}_V\left( X_A{}^M F_{MN} X^{NA}\right)+4\,{\rm Tr}_V\left( X^{MA} F_{MN} X_A{}^{N}\right),
\ee
where ${\rm Tr}(A) = A_E{}^E$. We see that this is the only part of $A$ whose derivative appears in the result. Note that we also expressed the action of the curvature on an object in ${\rm End}(V\otimes S_+)$ as a sum of two terms. In each of the last two terms in (\ref{off*}) both $F,X$ are viewed as objects in ${\rm End}(V\otimes S_+)$.

\subsection{Diagonal terms}

There is no derivatives of $A,\tilde{B}$ coming from the diagonal terms. Working their contributions is simply a tedious algebra. We start with the $E^2$ contributions. The upper diagonal term contributes
\be\label{d-3}
{\rm Tr}_V\left( - F_{MN} F^{MN} + 4( A_{MN} \tilde{B} +\tilde{B}A_{MN}) F^{MN} + 8{\rm Tr}(A^2) \tilde{B}^2 + 8 A_{M}{}^N \tilde{B} A_N{}^M \tilde{B} + 16 {\rm Tr}(A)\tilde{B}^3 + 8 \tilde{B}^4\right).
\ee
The lower-diagonal term gives
\be\label{d-4}
{\rm Tr}_V\left( - F_{M'N'} F^{M'N'} + 2{\rm Tr}(A^2){\rm Tr}(A^2)+4{\rm Tr}(A^2)({\rm Tr}(A)\tilde{B}+\tilde{B}{\rm Tr}(A)) + 4{\rm Tr}(A)\tilde{B}{\rm Tr}(A)\tilde{B}+ 4({\rm Tr}(A))^2\tilde{B}^2\right).
\ee

To work out the ${\cal F}^2$ contributions we first write down what one obtains for the maps on the upper-left and lower-right corners. We have
\be
\left[ \frac{1}{2} F_{MN} \epsilon_A{}^B - 2X_{A(M}X_{N)}{}^B \right]\epsilon_{M'N'} + \frac{1}{2} F_{M'N'}\epsilon_{NM}\epsilon_A{}^B
\ee
for the first diagonal element of the matrix of curvatures and
\be
\left[ \frac{1}{2} F_{MN} +X_{(M}{}^A X_{AN)} \right]\epsilon_{M'N'} \epsilon_{A'}{}^{B'} + \left[ \frac{1}{2} F_{M'N'}\epsilon_{A'}{}^{B'} - {\rm Tr}(X^2)\epsilon_{A'(M'} \epsilon_{N')}{}^{B'}\right] \epsilon_{NM}
\ee
for the second.

Squaring and taking the trace of the first gives
\be
{\rm Tr}_V\left( F_{MN}F^{MN} +F_{M'N'}F^{M'N'}- 4X^{MA} F_{MN} X_A{}^N + 4( X_{AM} X_N{}^B + X_{AN} X_M{}^B) X_B{}^M X^{NA} \right).
\ee
With a bit of algebra the last term here can be rewritten as follows
\be
{\rm Tr}_V\left( 4( X_{AM} X_N{}^B + X_{AN} X_M{}^B) X_B{}^M X^{NA}\right)={\rm Tr}_V\left( 4{\rm Tr}(X^4) - 8 {\rm Tr}(X^2) {\rm Tr}(X^2) \right).
\ee
Thus, overall, the square of the upper-left diagonal element gives
\be\label{d-1}
{\rm Tr}_V\Big(  F_{MN}F^{MN} +F_{M'N'}F^{M'N'}- 4X^{MA} F_{MN} X_A{}^N 
+ 4{\rm Tr}(X^4) - 8 {\rm Tr}(X^2) {\rm Tr}(X^2) \Big).
\ee

The trace of the square of the lower-right diagonal element gives 
\be
{\rm Tr}_V\left( F_{MN}F^{MN} + 4 X_A{}^M F_{MN} X^{NA} + 2(X_M{}^A X_{AN} + X_N{}^A X_{AM}) X^{MB} X_B{}^N\right) \\ \nonumber
+ {\rm Tr}_V\left( F_{M'N'}F^{M'N'} -6 {\rm Tr}(X^2){\rm Tr}(X^2)\right),
\ee
where the second line is from the square of the anti-self-dual $M'N'$ symmetric part. The last term in the first line can be brought into a more canonical form. Overall, after some algebra we get
\be\label{d-2}
{\rm Tr}_V\left( F_{MN}F^{MN} +F_{M'N'}F^{M'N'} + 4 X_A{}^M F_{MN} X^{NA} - 4{\rm Tr}(X^4) - 4 {\rm Tr}(X^2){\rm Tr}(X^2)\right).
\ee
The overall contribution from ${\cal F}^2$ is obtained by adding (\ref{d-1}) and (\ref{d-2}). We get
\be\label{d-5}
{\rm Tr}_V\left( 2F_{MN}F^{MN} +2F_{M'N'}F^{M'N'} - 4X^{MA} F_{MN} X_A{}^N + 4 X_A{}^M F_{MN} X^{NA} -12  {\rm Tr}(X^2){\rm Tr}(X^2)\right).
\ee
We can now substitute the matrix $X$. We have
\be
{\rm Tr}_V\left({\rm Tr}(X^2){\rm Tr}(X^2)\right) = {\rm Tr}_V\Big( {\rm Tr}(A^2){\rm Tr}(A^2)+ 2{\rm Tr}(A^2)({\rm Tr}(A)\tilde{B}+\tilde{B}{\rm Tr}(A)) \\ \nonumber + 2 {\rm Tr}(A)\tilde{B} {\rm Tr}(A)\tilde{B}+ 2({\rm Tr}(A))^2\tilde{B}^2+4{\rm Tr}(A^2) \tilde{B}^2 
+ 8{\rm Tr}(A)\tilde{B}^3 + 4\tilde{B}^4\Big).
\ee

Finally, adding (\ref{d-3}) and (\ref{d-4}), multiplying the result by $6$, and adding to (\ref{d-5}) we get
\be\label{d*}
{\rm Tr}_V\Big( -4F_{MN}F^{MN} -4F_{M'N'}F^{M'N'} + 24 ( A_{MN} \tilde{B} +\tilde{B}A_{MN}) F^{MN} 
\\ \nonumber
- 4X^{MA} F_{MN} X_A{}^N + 4 X_A{}^M F_{MN} X^{NA}+ 48 A_M{}^N \tilde{B} A_N{}^M \tilde{B} \Big),
\ee
with all other terms nicely cancelling away.

\subsection{The final result}

We can now combine the contribution (\ref{off*}) from off-diagonal terms with (\ref{d*}) to get the final result for the relevant heat kernel coefficient
\be\label{result}
a^{D^2}_4 = \frac{1}{(4\pi)^2} \frac{1}{12} \int {\rm Tr}_V \Big( 24 \, d_{AA'} {\rm Tr}(A) d^{AA'} \tilde{B} -4F_{MN}F^{MN} -4F_{M'N'}F^{M'N'} \\ \nonumber
+ 24 ( A_{MN} \tilde{B} +\tilde{B}A_{MN}) F^{MN}+ 48 A_M{}^N \tilde{B} A_N{}^M \tilde{B}  \Big).
\ee

\section{Calculation of the $\beta$-function}
\label{sec:beta}

With the hear-kernel coefficient (\ref{result}) at hand we can return to our problem of computing the $\beta$-function for (\ref{L-conn}). We now use (\ref{result}) with (\ref{A}), (\ref{tB}). 

\subsection{On-shell conditions}

The integrand in (\ref{result}) is not of the form (\ref{action}) of the action we started from. In particular, the first term in (\ref{result}) contains derivatives of the auxiliary field $B$. It is easy to see that this term goes away for the case of the usual YM theory, when the matrix $A_{AB}$ is a constant. However, for a general theory of the type we consider this derivative term is non-zero.

To eliminate the term in (\ref{result}) that contains unwanted derivatives, and also to be able to identify the $\beta$-function, we need to go on-shell and assume that the background satisfies its field equations. So, let us derive some consequences of the background field equations (\ref{feqs}). We already mentioned that the field equations can be viewed as giving rise to second order PDE's on the connection. However, we can also obtain some second order equations for the auxiliary field. Indeed, applying another derivative to the first equation in (\ref{feqs}) we get 
\be\label{on-shell}
0= d_A{}^{A'} d_{A'}{}^E B^a_{BE} = - \frac{1}{2} d^{MM'} d_{MM'} B^a_{BA} - \frac{1}{2} C^{aeb} F^e_A{}^E B^b_{BE}.
\ee
We note that the second term on the right-hand-side is automatically $AB$ symmetric as a consequence of the second equation in (\ref{feqs}). Indeed, the $AB$ anti-symmetric part is a multiple of
\be\label{AB-s}
C^{aeb} F^{e}_{MN} B^{b MN} = C^{aeb} (V')^{e}_{MN} B^{b MN} =0,
\ee
where the first equality is using the second of the field equations, and the last equality is a simple consequence of the gauge-invariance of $V(B)$. 

It is now clear that using the on-shell condition (\ref{on-shell}) we can replace the derivative term in (\ref{result}) by a term that contains no derivatives of $B$. To this end, it is convenient to write (\ref{on-shell}) as follows
\be\label{os-1}
0= d_A{}^{A'} d_{A'}{}^E \tilde{B}^{ab}_{BE} = - \frac{1}{2} d^{MM'} d_{MM'} \tilde{B}^{ab}_{BA} - \frac{1}{2} C^{aec} F^e_A{}^E \tilde{B}^{cb}_{BE} -  \frac{1}{2} C^{bec} F^e_A{}^E \tilde{B}^{ac}_{BE}.
\ee
Now, introducing the matrix $F^{ab}_{MN}\in {\rm End}(V)$
\be
F^{ab}_{MN}:= C^{aeb} F^e_{MN},
\ee
we can rewrite (\ref{os-1}) in the matrix form as
\be\label{os*}
- d^{MM'} d_{MM'} \tilde{B}_{BA} =  F_A{}^E\tilde{B}_{EB}- \tilde{B}_{EB} F_{A}{}^E.
\ee
 
\subsection{On-shell simplifications}

Let us do some rewriting of the first term in (\ref{result}). Substituting (\ref{os*}) we get
\be\label{fir-1}
-24 {\rm Tr}(A)^{(AB)}(F_A{}^E \tilde{B}_{EB} + \tilde{B}_B{}^E F_{EA}).
\ee
Note that we have added the symmetrisation on ${\rm Tr}(A)^{AB}$, which is for free because the quantity in the brackets is automatically $AB$ symmetric (as a consequence of field equations). 

Let us now consider the first term in the second line in (\ref{result}). It is given by
\be\label{sec-1}
24( A_{AM}{}^B{}_{N} \tilde{B}_B{}^A + \tilde{B}_A{}^B A_{BM}{}^A{}_N)F^{MN},
\ee
where all Lie algebra indices are suppressed and simple cyclic index contraction is assumed. We now assume that no gauge-fixing has been done on the $A$-matrix and it is symmetric in its pairs of indices. There is no harm in this assumption because the gauge-fixing part of $A$ anyway drops from the final result, as can be explicitly checked. The final trick is to do some exchanges of the spinor indices. For the first term this gives
\be
A_{AM}{}^B{}_{N} \tilde{B}_B{}^A = A_{NM}{}^B{}_{A} \tilde{B}_B{}^A + \epsilon_{AN} A_{EM}{}^{BE} \tilde{B}_B{}^A = \frac{1}{2} F_{MN} + {\rm Tr}(A)_M{}^B \tilde{B}_{BN},
\ee
where we have used the background field equation in the form
\be
\tilde{B}^{ab AB} A^{bc}_{ABCD} = -\frac{1}{2} F^{ac}_{CD}.
\ee
We have now spelled out the Lie algebra indices for clarity. For the second term we get similarly
\be
\tilde{B}_A{}^B A_{BM}{}^A{}_N = \tilde{B}_A{}^B A_{B}{}^A{}_{MN} + \epsilon_M{}^A \tilde{B}_A{}^B A_{BE}{}^E{}_N = \frac{1}{2} F_{MN} + \tilde{B}_{M}{}^B {\rm Tr}(A)_{BN}. 
\ee
Thus, overall, we can rewrite (\ref{sec-1}) as follows
\be\label{sec-2}
24 F_{AB} F^{AB} +24 {\rm Tr}(A)^{AB} (\tilde{B}_B{}^E F_{EA} +F_B{}^E \tilde{B}_{EA}),
\ee
where we have relabelled some indices. It is now clear that the first term (\ref{fir-1}) cancels some of the term just analysed. Indeed, we can further rewrite (\ref{sec-2}) as
\be\label{sec-3}
24 F_{AB} F^{AB} +24 {\rm Tr}(A)^{(AB)} (\tilde{B}_B{}^E F_{EA} +F_B{}^E \tilde{B}_{EA}) + 24 {\rm Tr}(A)_E{}^E \tilde{B}^{MN} F_{MN},
\ee
where we have used
\be
\tilde{B}^{MN} F_{MN} = F^{MN} \tilde{B}_{MN},
\ee
which is the already discussed consequence of the field equations. Now the second term in (\ref{sec-3}) cancels with (\ref{fir-1}). Thus, overall, on-shell the relevant heat-kernel coefficient becomes
\be\label{result*}
a^{D}_4 = \frac{1}{(4\pi)^2} \frac{1}{24} \int \left(8 F_{AB}F^{AB}  + 24 A_{AB}{}^{AB} \tilde{B}^{MN} F_{MN}+ 48 A_{AM}{}^{BN} \tilde{B}_{B}{}^C A_{CN}{}^{DM} \tilde{B}_D{}^A  \right),
\ee
where we divided the previous result by $2$ to obtain the coefficient for the original operator $D$.

We can now add the ghost contribution
\be
a_4^{ghost} = \frac{1}{(4\pi)^2} \frac{1}{12} \int F_{AB} F^{AB}
\ee
to get
\be\label{final*}
a^{D}_4 -2 a_4^{ghost} = \frac{1}{(4\pi)^2} \frac{1}{6} \int \left( F_{AB}F^{AB}  + 6 A_{AB}{}^{AB} \tilde{B}^{MN} F_{MN}+ 12 A_{AM}{}^{BN} \tilde{B}_{B}{}^C A_{CN}{}^{DM} \tilde{B}_D{}^A  \right),
\ee
which the final $\beta$-function quoted in (\ref{beta-f}). In the formula (\ref{beta-f}) we have also wrote $A,\tilde{B}$ explicitly in terms of the matrices of first and second derivatives of the function $f$. We note that this formula is only valid for the original matrix $A$ that is symmetric in its both index pairs. In other words, this is already a gauge-invariant formula from which the gauge-fixing parameters dropped out.

\section{Application to a 2-parameter family}
\label{sec:appl}

\subsection{2-parameter family of theories}

Let us consider a gauge theory containing just a single additional interaction term as compared to the standard YM
\be\label{L2}
{\cal L} = \frac{1}{4g^2} (F^a_{MN})^2 + \frac{\alpha}{3! g^2 M^2} C^{abc} F^a_A{}^B F^b_B{}^C F^c_C{}^A.
\ee
The sign in front of the first term is as appropriate for the Euclidean action. When the coupling constant is absorbed into the connection one obtains $\alpha g$ in front of the new term, which is a convenient parameterisation of the new interaction, as we shall see. It is our goal to obtain the $\beta$-function for $\alpha g$. 

To be able to use the result of the computation above, we first need to put (\ref{L2}) into the first order form. This is achieved by the following Lagrangian
\be\label{B2}
{\cal L} = B^{a MN} F^a_{MN} - g^2 (B^a_{MN})^2 + \frac{\alpha (2g^2)^2}{3M^2} C^{abc} B^a_A{}^B B^b_B{}^C B^c_C{}^A +O(B^4).
\ee
Indeed, varying with respect to $B$ we obtain
\be
F^a_{MN} = 2 g^2 B^a_{MN} + \frac{\alpha (2g^2)^2}{M^2} C^{abc} B_M^{b E} B^c_{EN} + O(B^3).
\ee
We can find $B^a_{MN}$ from here perturbatively, as a series expansion in $F$. We get
\be\label{B-F}
B^a_{MN} = \frac{1}{2g^2} F^a_{MN} - \frac{\alpha}{2g^2 M^2} C^{abc} F_M^{b E} F^c_{EN} + O(F^3).
\ee
Using the Jacobi identity we can rewrite this in the following form convenient for the latter
\be
\tilde{B}^{ab}_{MN} = \frac{1}{2g^2} F^{ab}_{MN}- \frac{\alpha}{g^2 M^2} F^{ae}_{(M}{}^E F^{eb}_{N)E}.
\ee
Substituting this into (\ref{B2}) we get (\ref{L2}), plus higher order corrections. 

\subsection{The matrix of second derivatives}

The matrix of second derivatives of the potential function is then as follows
\be
A^{ab}_{MANB} = - \frac{1}{2} (V'')^{ab}_{MANB} = -g^2 \delta^{ab} \epsilon_{M(N}\epsilon_{AB)} - \frac{2\alpha g^4}{M^2} C^{abc} ( \epsilon_{N(M} B^c_{A)B} +   \epsilon_{B(M} B^c_{A)N}).
\ee
Using spinor identities we can manipulate the last term into a more convenient form
\be
A^{ab}_{MANB} = - \frac{1}{2} g^2 \delta^{ab} (\epsilon_{MN}\epsilon_{AB}+\epsilon_{MB}\epsilon_{AN}) - \frac{2\alpha g^4}{M^2} ( \epsilon_{MN} \tilde{B}^{ab}_{AB} +   \epsilon_{AB} \tilde{B}^{ab}_{MN}),
\ee
where we have used the notation (\ref{tB}). The full trace is now easy to compute
\be
A^{ab}_{MA}{}^{MA} = -3g^2 \delta^{ab}.
\ee

\subsection{Computation}

Let us now substitute all the above into the final result (\ref{final*}). Using $C^{apq}C^{bpq}=C_2 \delta^{ab}$ the first term becomes $-C_2 (F^a_{MN})^2$. For the second term we get
\be\label{term-1}
6(-3)g^2\left( \frac{1}{2g^2} F^{ab}_{MN}- \frac{\alpha}{g^2 M^2} F^{ae}_{(M}{}^E F^{eb}_{N)E}\right) F^{ba MN} = 9C_2 (F^a_{MN})^2 + \frac{9\alpha C_2}{M^2} C^{abc} F^a_A{}^B F^b_B{}^C F^c_C{}^A,
\ee
where we used 
\be
C^{apb} C^{bqc} C^{cra} = - \frac{C_2}{2} C^{pqr}.
\ee
The computation of the last term in (\ref{final*}) is a bit harder, but the result is $1/3$ times (\ref{term-1}). Adding everything up we get
\be\label{2-p-2}
a_4^D- 2a_4^{ghost} = \frac{C_2}{(4\pi)^2} \frac{1}{6} \int 11 (F^a_{MN})^2 +  \frac{12 \alpha}{M^2} C^{abc} F^a_A{}^B F^b_B{}^C F^c_C{}^A.
\ee

\subsection{Interpretation}

The result (\ref{2-p-2}) directly gives the $\beta$-functions for the coefficients appearing in the Lagrangian (\ref{L2}). Indeed, we can immediate write
\be
\frac{\partial}{\partial \log{\mu}} \left(\frac{1}{4g^2}\right) = \frac{C_2}{(4\pi)^2} \frac{11}{6}, \qquad
\frac{\partial}{\partial \log{\mu}} \left(\frac{\alpha}{3! g^2M^2}\right) =  \frac{C_2}{(4\pi)^2 M^2} 2\alpha.
\ee
From here we get the familiar
\be
\frac{\partial g}{\partial \log{\mu}} = -\frac{11 g^3}{3} \frac{K}{(4\pi)^2},
\ee
which is unchanged in our deformed family of theories. The second equation gives, together with the $\beta$-function for the YM coupling
\be
\frac{\partial (\alpha g)}{\partial \log{\mu}} = \frac{ \alpha g^3 K}{(4\pi)^2},
\ee
which coincides with the result obtained in \cite{CFK} by a different method. 

\section{Discussion}

In \cite{CFK} we have motivated out interest in the family of theories (\ref{L-conn}) by a possibility that this family of theories may be renormalisable in the sense of effective field theory, i.e. after all possible field redefinitions are taken into account. The main result of this paper is that this is indeed the case at the one-loop level. We have also explicitly computed the arising renormalisation group flow (\ref{beta-f}). 

Perhaps a good analogy indicating why our result (\ref{beta-f}) is interesting is to the non-linear $\sigma$-model in $2+\epsilon$ dimensions. This is a model with the Lagrangian
\be\label{sigma}
S[\phi]=\int g_{ij}(\phi(x)) \partial^\mu \phi^i(x) \partial_\mu \phi^j(x),
\ee
where $\phi^i(x)$ is a map from a two-dimensional surface to some target space with metric $g_{ij}$. Because in 2 dimensions the field $\phi^i$ is dimensionless, all powers of the field are allowed as renormalisable interactions, and so the above field theory is renormalisable. It is a famous result \cite{Friedan:1980jf} that the $\beta$-function for the infinite number of coupling constants stored in the metric $g_{ij}(\phi)$ is given by the Ricci tensor $R_{ij}$, so that the renormalisation group flow is the Ricci flow. 

In this paper we have computed the one-loop renormalisation group flow (\ref{beta-f}) for theory (\ref{L-conn}). The result is a flow in the space of scalar functions of the self-dual part of the field strength, and is in a sense an analog of the Ricci flow for (\ref{sigma}). It would be very interesting to understand a geometrical interpretation of (\ref{beta-f}), if any, but we leave this to future work. 

The most important open question is whether the renormalisability continues to hold at higher loops, or is lost at some sufficiently high loop order. Indeed, any field theory is renormalisable once all terms compatible with the symmetries are included into the Lagrangian. This is renormalisability in the sense of effective field theory. However, the effective field theory Lagrangians with their infinite number of terms of ever increasing number of derivatives are too complicated to explicitly compute the associated renormalisation group flow. We are guided by a vision that there may exist four-dimensional renormalisable effective field theory models that come with an infinite number of couplings, but that are still sufficiently simple so that explicit calculations of the renormalisation group flow in the infinite-dimensional space of couplings are possible. The class of theories studied in this paper is an attempt to construct such a model. 

Our result that the theories (\ref{L-conn}) are one-loop renormalisable says little about what is to happen at higher loops. Indeed, there are examples of power-counting non-renormalisable theories that are renormalisable at one loop, but this property is lost at higher loops. One such example is General Relativity. Indeed, it is a famous result \cite{'tHooft:1974bx} that GR is one-loop finite with zero $\Lambda$, and one-loop renormalisable with $\Lambda\not=0$. At two loops this property is lost \cite{Goroff:1985th}. In the case of GR there is a single coupling, and so perhaps this is not a good analogy with the infinite-parametric models considered here.  Another, more appropriate example is that of scalar field in 4 dimensions with an arbitrary potential $V(\phi)$. It is an easy exercise to check that this theory is one-loop renormalisable, with the one-loop $\beta$-function for $V(\phi)$ being a multiple of $(V'')^2$. However, this property is lost at two loops. 

In the case of the theories considered here, our justification for the renormalisability hopes is the "collapsing" property of perturbation theory explained in details in \cite{CFK}. This property implies that there is at most a single derivative on internal Feynman diagram lines, and leads to improved UV behaviour. In the considered in this paper first derivative formulation this single derivative property is realised automatically, because in this formulation the vertices contain no derivatives, and the propagator is the inverse of a first-order operator, which thus can be written as a single derivative of an inverse of a second-order operator. Thus, it is again the "collapsing" property of our model that allowed us to perform the one-loop computation and obtain an analog of the Ricci flow for (\ref{sigma}), but now in the context of a power-counting non-renormalisable theory. We hope that this property will also make the higher loop behaviour nice (renormalisable), but at present we have no calculations to support these hopes.  

The question of whether the renormalisability continues to hold at higher loops is related to the  question whether and in what regime one can trust the flow (\ref{beta-f}). Certainly, even if a model is renormalisable, the one-loop flow can only be trusted if higher loop corrections are negligible, i.e. in the regime of validity of the perturbation theory. So, if our model continues to be renormalisable at all loops, we would just need to compute (or estimate) the associated $\beta$-function corrections to see in what regime the one-loop flow can be trusted. However, in case renormalisability is lost at some loop level, one needs to add new operators to the tree level Lagrangian. The presence of these would affect the one-loop flow, as well as the very ability to compute such a flow. So, at present we do not understand how to establish the regime of validity (if any) of a one-loop flow for a theory that becomes non-renormalisable at some higher loop order. And this is why hope, possibly over-optimistically, that there are models of the type considered here that exhibit all-loop renormalisability and allow for an explicit computation of the arising renormalisation group flow. 

If the flow (\ref{beta-f}) can be trusted, at least in some regions in the space of couplings, the most interesting question is where this flow takes one in the ulta-violet. The first term in (\ref{beta-f}) is non-positive, while the last term is non-negative. Thus, there can in principle exist functions $f$ which give the vanishing $\beta$-function. These can be IR or UV fixed points. Since the theory at such a fixed point must be scale free, the corresponding function $f$ must be a homogeneous degree 2 function of its argument, so that the dependence on $M$ drops out. However, given that the number of arguments of the function $f$ is in general as large as $2\,{\rm dim}(G)-3$, there is a vast range of possibilities even if the requirement of homogeneity is imposed. At present we are unable to say anything about the existence of fixed points other than the one corresponding to free theory that is realised in the usual YM theory. 

If there are non-trivial fixed points, the next question is about the dimension of the critical surfaces. For a given fixed point this is the surface of all the trajectories that lead to it. If there are fixed points with a finite-dimensional critical surface, one has a realisation of the asymptotic safety scenario of Weinberg, see \cite{Weinberg:2009bg} for a recent discussion. What is most interesting about our result (\ref{beta-f}) is that it now allows for a very concrete study of these questions, for the first time in the four-dimensional setup. 

It is pertinent to compare our approach to that followed in the modern asymptotic safety literature, see e.g. \cite{Percacci:2007sz}. In this approach, a version \cite{Wetterich:1992yh} of Polchinski's exact renormalisation group equation \cite{Polchinski:1983gv} is used to compute and study the flow. To compute anything in practice, a truncation in the space of effective actions must be taken, and then one studies this truncated flow. In the case of our model, because of renormalisability, no truncation is necessary, at least in the presently studied one-loop setup. Another important difference between this work and the approach reviewed in \cite{Percacci:2007sz} is that in the latter divergences other than logarithmic are taken seriously, and one studies the flow for dimensionful couplings. In contrast, in this work we used the dimensional regularisation, kept only the logarithmic divergences and obtained the flow for dimensionless couplings. The $\beta$-function we computed is independent of the gauge-fixings used at intermediate stages (this has been used as a check of the final result). What we obtained is a flow in the space of observable coupling constants, with the latter being operationally defined from e.g. on-shell gluon scattering amplitudes. 

Let us further remark that our result (\ref{beta-f}) may be interesting even if the hopes about all-loop renormalisability are too optimistic. Another possible scenario could be that there are some smaller, perhaps finite dimensional, subspaces in the space of theories of type (\ref{L-conn}) such that if one is inside the subspace the renormalisation does not take one out. Indeed, the usual YM theory is precisely such a one-dimensional subspace. It is possible that one can find other subspaces of this type by studying the flow (\ref{beta-f}). This would be interesting, and would give new gauge theories in four space-time dimensions. 

The results of this paper also give support to the conjecture stated in \cite{Krasnov:2006du} that a certain infinite-parametric class of four-dimensional gravity theories is (one-loop) renormalisable. In fact, the structures that arise in the context of these gravity theories are quite analogous to what we have encountered here. Similarly to what we have seen for the case of gauge theories (\ref{L-conn}), a straightforward application of the background field method for gravity theories  \cite{Krasnov:2006du} leads to a non-Laplace type operator, see \cite{Groh:2013oaa}. One lesson from the calculation performed in this paper is that an efficient strategy for dealing with such operators is via the first order formalism. Work on applying these ideas to gravity is currently in progress.

\section*{Acknowledgements} The author was supported by ERC Starting Grant 277570-DIGT. I am grateful to Dima Vassilevich for discussions on the subject of this calculation, and in particular for the suggestion to square the first order operator, which proved so useful. I am also grateful to Laurent Freidel for comments on the manuscript.

\end{document}